\documentclass[preprint,amsmath,amssymb,prb,longbibliography]{revtex4-1}

\usepackage{graphicx}  
\usepackage{dcolumn}   
\usepackage{bm}   
\usepackage[usenames,dvipsnames]{xcolor}
\usepackage[colorlinks=true,linkcolor=blue,citecolor=blue,urlcolor=blue]{hyperref}
\def\ev#1{\langle #1 \rangle}
\newcommand{\iu}{\mathrm{i}}

\begin{document}

\title{\textit{Ab initio} investigation of impurity-induced in-gap states in 
Bi$_2$Te$_3$ and Bi$_2$Se$_3$} 

\author{Juba Bouaziz}
\email{j.bouaziz@fz-juelich.de}
\author{Manuel dos Santos Dias}
\email{m.dos.santos.dias@fz-juelich.de}
\author{Julen Iba\~nez-Azpiroz}
\author{Samir Lounis}
\email{s.lounis@fz-juelich.de}
\affiliation{Peter Gr\"unberg Institut and Institute for Advanced Simulation, 
Forschungszentrum J\"ulich and JARA, 52425 J\"ulich, Germany}

\date{\today}

\begin{abstract}
We investigate in-gap states emerging when a single $3d$ transition metal impurity is 
embedded in topological insulators (Bi$_2$Te$_3$ and Bi$_2$Se$_3$). We use a combined 
approach relying on first-principles calculations and an Anderson impurity model.
By computing the local density of states of Cr, Mn, Fe and Co embedded not only in surfaces of Bi$_2$Te$_3$ 
and of Bi$_2$Se$_3$ but also in their bulk phases, we demonstrate that in-gap states originate from the 
hybridization of the electronic states of the impurity with bulk bands and 
not with the topological surface states as it is usually assumed. This finding is analyzed using 
a simplified Anderson impurity model. These 
observations are in contradiction with the prevailing models used to investigate the 
magnetic doping of topological insulators [R. R. Biswas \textit{et al.} PRB \textbf{81}, 
233405 (2010), A. M. Black-Schafer \textit{et al.} PRB \textbf{91}, 201411 (2015)], 
which attribute the origin of the in-gap states to the hybridization with the topological 
surface states. 
\end{abstract}

\maketitle
\section{Introduction}
The concept of band topology was introduced in condensed matter physics in the 
context of the quantum Hall effect (QHE)~\cite{Klitzing:1980,Vonkiltzing:1986}, 
which represents a quantized version of the classical Hall effect~\cite{Hall:1879}. 
The QHE is observed in two-dimensional electronic systems at low temperatures 
and under strong magnetic fields applied perpendicularly to the plane containing 
the electrons~\cite{Klitzing:1980,Vonkiltzing:1986}. In the classical picture, 
the electrons can be viewed as charges moving in circles around a magnetic field. 
At the edges of the sample, the circles are not completed giving rise to chiral 
edge states~\cite{Halperin:1982}. A similar scenario occurs in two dimensional 
topological insulators without external magnetic fields, \textit{i.e.}~no time-reversal 
symmetry (TRS) breaking. This mechanism is induced by the intrinsic spin-orbit 
interaction (SOI)~\cite{Zhang:2009,Hasan:2010,Qi:2011} acting as an effective 
magnetic field $\vec{B}_\text{eff}(\vec{k})$, which couples to the electron spin. 
$\vec{B}_\text{eff}(\vec{k})$ is an odd function of $\vec{k}$ due to time reversal 
symmetry, thus, electrons moving in opposite directions along the edges have 
opposite spin polarizations. This is known as the quantum spin Hall effect 
(QSHE)~\cite{Kane:2005,Fu:2007,Konig:2007}, where the edge states are topologically 
protected from backscattering due to spin-momentum locking and TRS~\cite{Lang:2007,Hasan:2010}.
The topological protection results in a dissipationless transport even in presence 
of non-magnetic disorder~\cite{Hasan:2010}. In three dimensions (3D), topological 
insulators are insulating bulk materials but possess metallic edges states located 
at the surface. The first 3D topological insulator identified experimentally was 
Bi$_{1-x}$Sb$_{x}$~\cite{Hsieh:2008}. Afterwards, a multitude of 3D topological 
insulators were discovered, such as Bi$_2$Se$_3$, Bi$_2$Te$_3$ and 
Sb$_2$Te$_3$~\cite{Zhang:2009,Chen:2010,Hor:2009}. As a characteristic feature, 
the band structure of these materials displays a topological surface state that 
has a dispersion linear in $\vec{k}$ near the $\Gamma$ point~\cite{Zhang:2009}.

The TRS in topological insulators can be broken either by using an external magnetic 
field or by doping the material with magnetic impurities, which can generate numerous 
interesting effects. One of the most remarkable is the quantum anomalous Hall effect 
(QAHE), whereby the measured conductance is quantized to integer multiples of $e^{2}/h$,
the integer being the Chern number of the system~\cite{annurev-conmatphys-031115-011417}.
This effect has great potential for future devices with low power consumption
that rely exclusively on the electron spin~\cite{Garate:2010,Fu_Liang_2:2009,Tse:2010}.
Interestingly, the doping of Bi(Sb)$_2$Te$_3$ and (Bi$_{1-x}$Sb$_{x}$)$_{2}$Te$_3$ with 
high concentrations of Cr magnetic impurities allowed the experimental realization of 
the QAHE at low temperatures~\cite{Chang:2013,Kou:2014,Checkelsky:2014}.
The observation of the QAHE is a signature of a gap 
opening at the Dirac point. However, this gap opening is still a controversial point, which 
generated several experimental investigations on  magnetically doped topological 
insulators. Chen \textit{et} \textit{al}.~\cite{Chen:2010} used angle-resolved photoemission 
(ARPES) and showed that in non-magnetically doped Bi$_2$Se$_3$ no gap opening is 
observed, while the presence of Fe impurities breaking TRS leads to a gap opening at 
the $\Gamma$ point, which was attributed to the presence of a ferromagnetic order of 
the magnetic impurities. Similar observations concerning the gap opening were made 
later in Refs.~\onlinecite{Wray:2011,Xu:2012,Lee:2015,Barringa:2016}. However, other 
experimental works reported the absence of a gap. Using ARPES, Scholz \textit{et} 
\textit{al}.~\cite{Scholz:2013} showed that Fe impurities deposited on a Bi$_2$Se$_3$ 
surface at high or low temperatures do not lead to a gap opening at the Dirac point. 
In fact, the surface state remains robust even for high impurity coverage. Afterwards, 
other works~\cite{Honolka:2012,Sck:2013} combining various experimental methods (ARPES 
or scanning tunneling microscopy) with \textit{ab initio} calculations did not report 
a gap opening at the Dirac point. 

The existence of long range ferromagnetic order and the absence of a gap opening
can be reconciled when impurity-induced in-gap states are present. 
Indeed, based on scanning tunneling spectroscopy (STS) measurements and 
a phenomenological model for the topological surface state, Sessi \textit{et al.}~\onlinecite{Sessi:2016} 
showed that in-gap states lead to a local filling of the band gap in presence of 
ferromagnetic order. These in-gap states consist of sharp resonances in the density 
of states lying within the bulk band gap. Interestingly, similar resonances were also observed 
from experiment and first-principles calculations for $3d$ impurities deposited 
on a Cu(111) surface~\cite{Limot:2005,Samir:2006}. The resonances were located at 
the bottom of the surface state with a broadening caused by the hybridization with 
the bulk bands. The presence of these peaks is explained by the nature of the impurity 
potential, and its strength, which can attract or repel electronic states of the 
host material.

In topological insulators, the in-gap states have been investigated with 
phenomenological models~\cite{Biswas:2010,Black-Schaffer:2015}, which relate their 
creation to the presence of topological surface states. The latter are modeled 
using a Dirac-like Hamiltonian with a linear dispersion~\cite{Black-Schaffer:2015}:
\begin{equation}
\boldsymbol{H}_\text{D} = \hbar\,v_\text{F}\,\big(\vec{\boldsymbol{\sigma}}
\times\vec{k}\big)_{z}\quad,
\label{Linear_Dirac_Hamiltonian}
\end{equation}
with $v_\text{F}$ being the Fermi velocity and $\vec{k}$ the electron momentum.
The magnetic impurities are modeled by a scattering potential $\boldsymbol{V}(\vec{r})$, 
which contains a non-magnetic and a magnetic part: $\boldsymbol{V}(\vec{r}) = 
\big(U\,\boldsymbol{\sigma}_{0} - J\,\vec{M}\cdot\vec{\boldsymbol{\sigma}}\big)
\,\delta(\vec{r}-\vec{r}_{n})$, $\boldsymbol{\sigma}_{0}$ is the $2\times2$ identity matrix.
The spatial dependence of the potential is approximated by a delta function 
and $\vec{r}_{n}$ is the position of the impurity. $U$ represents the strength 
of the charge scattering, while $J$ is the coupling constant between the spin 
of the surface electrons and the magnetic moment $\vec{M}$ of the impurity.
In presence of a weak charge scattering (\textit{i.e.}~$\frac{U}{JM}<1$) and 
when considering $\vec{M}\parallel z$, the energy dispersion is gapped near 
the Dirac point~\cite{Black-Schaffer:2015,Liu:2009}. However, magnetic 
impurities can induce a strong charge scattering, which leads to the creation 
of in-gap states filling locally the band gap. This phenomenological model 
seems to provide several possible outcomes susceptible to explain the experimental 
observations. Nonetheless, it does not have a full description of the electronic 
structure and misses the contribution of the bulk bands, which may play a 
crucial role in the formation of these in-gap states.

In this work, we explore the impact of the bulk bands on the in-gap states 
present in magnetically doped topological insulators~\cite{Chotorlishvili:2014,Antonov:2017,Sessi:2016}. 
We performed first-principles  calculations for $3d$ transition metal 
impurities embedded in the surfaces of Bi$_{2}$Te$_{3}$ and Bi$_{2}$Se$_{3}$, 
fixing $\vec{M}\parallel z$. Our calculations show that the band gap is locally 
filled by impurity resonances, as predicted in Refs.~\onlinecite{Biswas:2010,Black-Schaffer:2015,Sessi:2016}.
Comparing with calculations for $3d$ impurities embedded in the bulk of 
Bi$_{2}$Te$_{3}$ and Bi$_{2}$Se$_{3}$, we demonstrate that the hybridization 
of the impurity $d$-states with the $sp$ bulk bands is the driving mechanism 
behind the formation of these in-gap states. A simple model is used to analyze 
our results.

This paper is organized as follows. In Sec.~\ref{method_description}, we describe 
the first-principles approach and provide the technical details used in the 
calculations. We also consider a simple Anderson model, which will be employed 
to understand the emergence of in-gap states in magnetically doped topological 
insulators. Sec.~\ref{Results_in_gap_states} is dedicated to the study of the 
in-gap states from first-principles by highlighting the importance of the contribution 
of bulk states. Furthermore, we show the results of the local density of states 
obtained within the Anderson model and explore the possibility of creating in-gap 
states from hybridization with the bulk or the topological surface state. 
Finally, in Sec.~\ref{conclusions_in_gap_states} we summarize our results.

\section{Description of the methods and computational aspects}
\label{method_description}

\subsection{\textit{Ab initio} method}
Our first-principles simulations are performed using the Korringa-Kohn-Rostoker Green 
function (KKR-GF) method~\cite{Bauer:2014,Papanikolaou:2002}, which relies on multiple 
scattering theory. The calculations are carried out using the atomic sphere approximation 
(ASA) including full charge density in the local spin density approximation (LSDA), as 
parametrized by Vosko, Wilk, and Nusair~\cite{Vosko:1980}. We employ the scalar relativistic 
approach augmented self-consistently with SOI, which is of crucial importance in topological 
insulators. The calculations are made in two steps. First, we simulate the periodic 
host (bulk and surface), which is then used to self-consistently embed the impurities 
in real space. 

\begin{figure}[b]
  \includegraphics[width=\columnwidth]{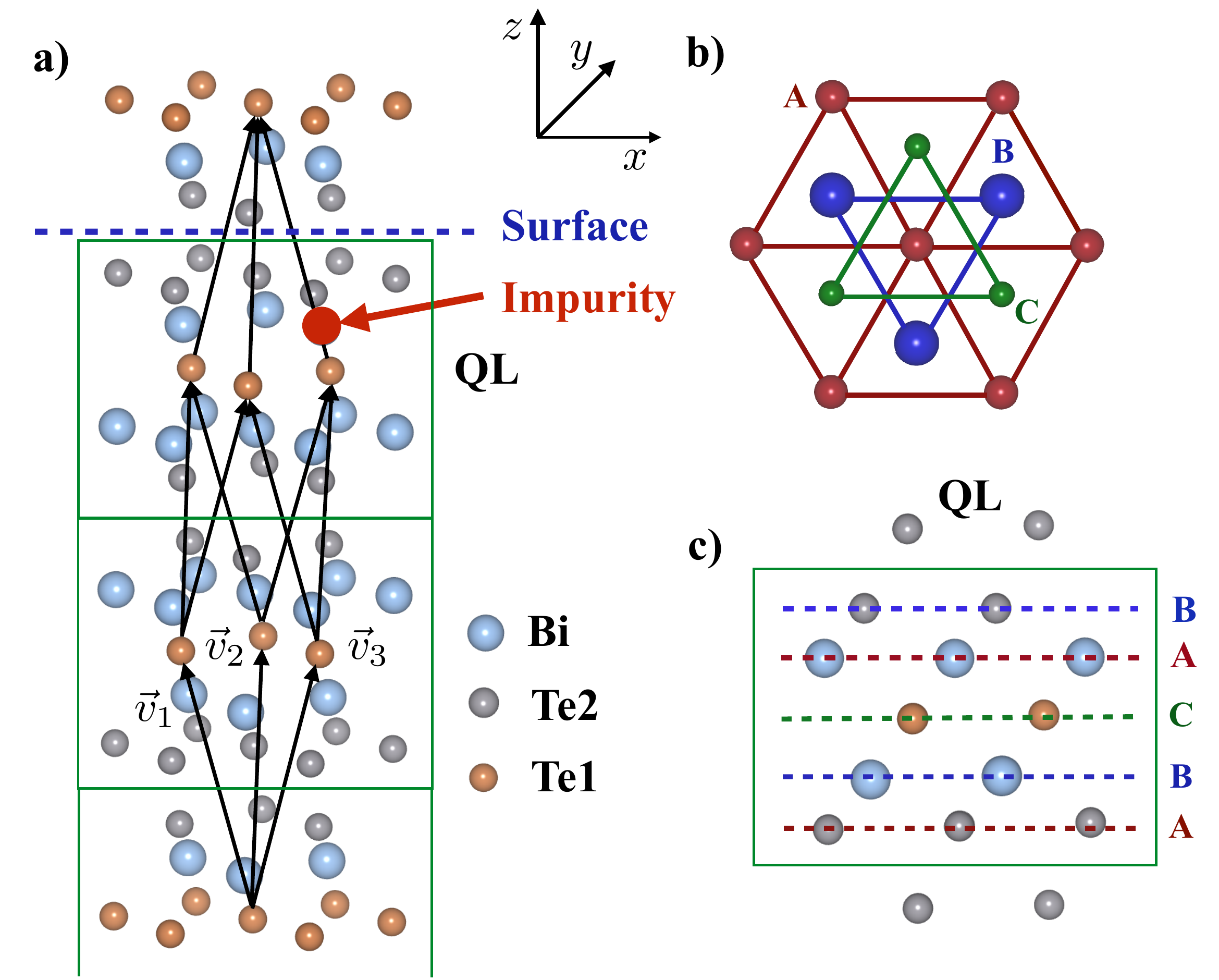}
  \caption{\label{Bi2Te3_crystal_structure}
  Atomic structure.
  (a) Crystal structure of Bi$_{2}$Te$_{3}$ characterized by a primitive unit cell 
  containing five inequivalent atoms (two Bi, two Te2 and one Te1). The rhombohedral 
  primitive vectors $\{\vec{v}_{1},\vec{v}_{2},\vec{v}_{3}\}$ are also shown. For the 
  slab, six quintuple layers are used. The cut to create the surface is indicated on 
  the figure. The position of the impurity in the real space calculations is indicated 
  by a red arrow. (b) Top view of the crystal structure showing the ABC stacking of the 
  different layers. (c) Side view of the quintuple layer showing the stacking of the 
  layers along the $z$-direction.}
\end{figure}

The bulk unit cell includes five atoms and ten sites in total (\textit{i.e.}~five vacuum sites), 
see Fig.~\ref{Bi2Te3_crystal_structure}(a). The lattice parameters for the rhombohedral crystal 
structure are taken from experiment~\cite{Wei:2010}. The system is self-consistently converged 
using $30 \times 30 \times 30$ $k$-points in the full Brillouin zone. For the surface calculations, 
we used a slab containing six quintuple layers, which is enough to ensure the absence of a direct 
coupling between the two surfaces, since the surface state penetrates only within two quintuple 
layers~\cite{Wei:2010}. The slab was converged using a $60 \times 60 \times 1$ $k$-mesh in the full 
Brillouin zone. Once the Green functions of the periodic system are obtained, the impurity is embedded 
self-consistently in a finite region in space (\textit{i.e.}~the perturbation in the potential 
due to the impurity has a finite range) using the following Dyson equation~\cite{Gonis:2000,Zabloudil:2005}:
\begin{equation}
\begin{split}
{G}_\text{I}(\varepsilon) & =  {G}_\text{H}(\varepsilon) + {G}_\text{H}(\varepsilon)\,
({V}_\text{I}-{V}_\text{H})\,{G}_\text{I}(\varepsilon)\quad,\\ 
& = {G}_\text{H}(\varepsilon) +{G}_\text{H}(\varepsilon)
\,\Delta{V}\,{G}_\text{I}(\varepsilon)\quad,
\end{split}
\label{Impurity_dyson_equation}
\end{equation}
where the real space dependence is omitted for simplicity.
${V}_\text{I}$ (${V}_\text{H}$) is the Kohn-Sham potential in the cluster in presence (absence) of the impurity.
${G}_\text{I}(\varepsilon)$ represents the Green function of the system in presence of the impurity, while ${G}_\text{H}(\varepsilon)$ is the Green function of the undisturbed host.
The $3d$ impurities are substituting a Bi atom from the subsurface layer as indicated in Fig.~\ref{Bi2Te3_crystal_structure}(a).
This position is stable thermodynamically as observed experimentally and predicted from first-principles for the case of Fe in Bi$_2$Te$_3$~\cite{West:2012,Abdalla:2013}.
We considered different real space cluster sizes and found that the quantities of interest (local density of states, spin 
and orbital moments) are converged when considering a cluster containing 102 
sites in total ($24$ Bi atoms, $31$ Te (Se) atoms and $47$ vacuum sites).

\subsection{Anderson model}
\label{Anderson_model_theory}
In order to explain the origin of the in-gap states observed in the local density of states (LDOS) of the $3d$ impurities 
embedded in Bi$_2$Te$_3$ (Bi$_2$Se$_3$), we use a simple Anderson impurity model.
Our goal is to understand if and how the hybridization with both bulk and surface states 
leads to the formation of in-gap states. We consider a minimal model to interpret our 
first-principles calculations, which takes into account an impurity with a single $d$-orbital 
(as an example, we chose the $d_{xy}$ component). The impurity has two energy levels 
$\varepsilon^{\uparrow}_{i}$ and $\varepsilon^{\downarrow}_{i}$ for the majority and 
minority spin channels, respectively. This impurity hybridizes with a two-dimensional 
topological insulator surface state, which is characterized by the linear Dirac Hamiltonian 
shown in Eq.~\eqref{Linear_Dirac_Hamiltonian}. Furthermore, we account for the bulk bands 
by including Bloch states which are also characterized by the wave vector $\vec{k}$ and 
the eigenenergies $\varepsilon^{\sigma}_{\vec{k}}$ for each spin channel. The bulk and 
surface states interact only via the impurity. The impurity LDOS can be computed using 
the local Green function for which the spin diagonal part reads: 
\begin{equation}
\boldsymbol{G}^{\sigma\sigma}_{i}(\varepsilon) = \frac{1}{\varepsilon - \varepsilon^{\sigma}_{i} - 
	\Sigma(\varepsilon)}\quad,
\label{Gf_imp_tss_bulk}
\end{equation}
$\Sigma(\varepsilon)$ represents the hybridization function of the impurity. It 
describes the hybridization between the impurity $d$-electrons and the host $sp$-electrons. 
$\Sigma(\varepsilon) = \Lambda(\varepsilon) + \iu\Delta(\varepsilon)$, with
$\Lambda(\varepsilon)$ and $\Delta(\varepsilon)$ being the real and imaginary parts 
of $\Sigma(\varepsilon)$, respectively. 
For the purpose of our study, we consider the self-energy to be spin-independent.
The spin polarized LDOS of the impurity 
$n^{\sigma}_{i}(\varepsilon)$ is given by: 
\begin{equation}
n^{\sigma}_{i}(\varepsilon) = -\frac{1}{\pi} \frac{\Delta(\varepsilon)}{(\varepsilon - 
\varepsilon^{\sigma}_{i}-\Lambda(\varepsilon))^2 + (\Delta(\varepsilon))^2}\quad.
\label{ldos_model_anderson}
\end{equation}
$\Sigma(\varepsilon)$ can also be decomposed into bulk and surface contributions: 
\begin{equation}
\Sigma(\varepsilon) = \Sigma^\text{b}(\varepsilon) + \Sigma^\text{s}(\varepsilon)\quad,
\end{equation}
where $\Sigma^\text{b}(\varepsilon)$ and ${\Sigma}^\text{s}(\varepsilon)$ represent 
the hybridization functions with the bulk and surface states, respectively. 
Moreover, we define $\Delta_\text{b}(\varepsilon)$ $(\Delta_\text{s}(\varepsilon))$ 
as the imaginary part of the bulk (surface) hybridization function, while 
$\Lambda_\text{b}(\varepsilon)$ $(\Lambda_\text{s}(\varepsilon))$ is the real 
part of the  bulk (surface) hybridization function. Relying on Eq.~\eqref{ldos_model_anderson}, 
we expect in-gap states to occur when $\varepsilon - \varepsilon^{\sigma}_{i}-\Lambda_\text{b}(\varepsilon)-\Lambda_\text{s}(\varepsilon) \simeq 0$ and for $(\Delta_\text{b}(\varepsilon)+\Delta_\text{s}(\varepsilon))$ small. The LDOS of the 
impurities is mildly affected when they are moved from the surface to the bulk as 
shown in Fig.~\ref{LDOS_3d_Bi2Te3} and discussed in the next sections. Therefore, one can conclude that the coupling 
to the surface state is rather weak. $\Sigma^\text{b}_{i}(\varepsilon)$ and ${\Sigma}^\text{s}_{i}(\varepsilon)$ are 
derived analytically in   Appendix~\ref{Append_A} under the following assumptions: 
on the one hand, the bulk band is modeled using a gapped density of states $n_\text{b}(\varepsilon)$ 
given by: 
\begin{equation}
n_\text{b}(\varepsilon) =
\left \{
\begin{tabular}{ccc}
$n_\text{b}$ \quad {for} \quad $\varepsilon_\text{bv}<\varepsilon<\varepsilon_\text{tv}$\quad, \\
$n_\text{b}$ \quad {for} \quad $\varepsilon_\text{bc}<\varepsilon<\varepsilon_\text{tc}$\quad,   \\
0   \quad {elsewhere} \quad.
\end{tabular}
\right.
\label{bulk_ldos_and}
\end{equation}
$n_\text{b}$ is the occupation number in the valence and conduction band. $\varepsilon_\text{bv}$ $(\varepsilon_\text{bc})$ and $\varepsilon_\text{tv}$ $(\varepsilon_\text{tc})$ represent the bottom and the top of the valence (conduction) band, respectively.
On the other hand, the surface state is described by the Dirac Hamiltonian given in Eq.~\eqref{Linear_Dirac_Hamiltonian}. The LDOS of the surface states is linear within a certain energy window and is then connected to a flat LDOS, which is required to avoid spurious peaks in the LDOS at the cutoff energies $\pm\varepsilon_{0}$: 
\begin{equation}
n_\text{s}(\varepsilon) =
\left \{
\begin{tabular}{ccc}
$\frac{|\varepsilon-\varepsilon_\text{D}|}{\varepsilon^{2}_{0}}$ \quad {for} \quad $-\varepsilon_0<\varepsilon<\varepsilon_0$\quad,\\
$\frac{1}{\varepsilon_{0}}$ \quad {for} \quad $\varepsilon_\text{tv}<\varepsilon<-\varepsilon_{0}$\quad,\\
$\frac{1}{\varepsilon_{0}}$ \quad {for} \quad $\varepsilon_0<\varepsilon<\varepsilon_\text{tc}$\quad,\\
0   \quad {elsewhere} \quad.
\end{tabular}
\right.
\end{equation}
$\varepsilon_\text{D}$ being the energy of the Dirac point with respect to the Fermi energy (chosen to be the reference energy).
This model can be parametrized using first-principles data, 
which we proceed to discuss next.

\section{Results and discussion}
\label{Results_in_gap_states}
\subsection{In-gap states from first-principles}

\begin{table}[b]
\begin{center}
\begin{tabular}{lccc}
\hline
Element    & $Q$  &  $M_\text{s} (\mu_\text{B})$ & $M_\text{l} (\mu_\text{B})$ \\
\hline
Cr (Bi$_2$Te$_3$)     & 5.154  &  3.843  &  0.065 \\
Cr (Bi$_2$Se$_3$)     & 4.841  &  3.671  &  0.008  \\
\hline
Mn (Bi$_2$Te$_3$)    & 6.160  &  4.412  &  0.050 \\
Mn (Bi$_2$Se$_3$)    & 5.863  &  4.421  &  0.024  \\
\hline
Fe  (Bi$_2$Te$_3$)    & 7.282  &  3.395  &  0.260 \\
Fe  (Bi$_2$Se$_3$)    & 6.963  &  3.482  &  0.144  \\
\hline
Co  (Bi$_2$Te$_3$)   & 8.448  &  2.108  &  0.883 \\
Co  (Bi$_2$Se$_3$)   & 8.136  &  2.231  &  0.942  \\
\hline
\end{tabular}
\caption{Ground state properties of $3d$ impurities embedded in the Bi$_2$Te$_3$ and Bi$_2$Se$_3$ (111) surfaces: valence charge on the impurity $Q$, spin moment $M_\text{s}$ and orbital moment $M_\text{l}$.}
\label{3d_Bi2Te3_gs}
\end{center}
\end{table}

We show in Table~\ref{3d_Bi2Te3_gs} the charge, spin and orbital moments of single $3d$ 
transition metal impurities: Cr,  Mn, Fe and Co, which are embedded into the 
Bi$_2$Te$_3$ and Bi$_2$Se$_3$ surfaces. The spin moments are considered to be normal to each surface.
The valence charge on the impurity is shown in the first column. 
We notice that all $3d$ impurities are donors of electrons (n-type doping).
Similar results were obtained for Fe impurities in Bi$_2$Te$_3$~\cite{West:2012}.
The second column in Table.~\ref{3d_Bi2Te3_gs} displays the values of the spin moment $M_\text{s}$.
Cr and Mn have a nearly half-filled $d$-shell and present high values for $M_\text{s}$, which decreases for Fe and Co following the Hund's rules. The values of the orbital moments ($M_\text{l}$) shown 
in the third column behave differently. High values are obtained for Fe and Co due to the partial 
filling of the minority $d$-orbitals, in contrast to the low values of Cr and Mn. 
The impact of the chemical nature of the substrate was also investigated. When $3d$ impurities 
are embedded in the surface of Bi$_2$Se$_3$ instead of Bi$_2$Te$_3$, the following 
changes occur in the ground state quantities: First, the impurities tend to donate 
more electrons. Second, $M_\text{s}$ decreases for Cr while it increases for Mn, Fe 
and Co. Finally, $M_\text{l}$ is substantially affected, largely decreasing in the 
case of Cr, Mn and Fe, while increasing for Co (see Table.~\ref{3d_Bi2Te3_gs}). 
This large effect of the
substrate on the magnitude of $M_\text{l}$ is due to the high sensitivity
of this quantity to the details 
of the hybridization.

\begin{figure*}
	\centering
	\includegraphics[width=1.0\textwidth]{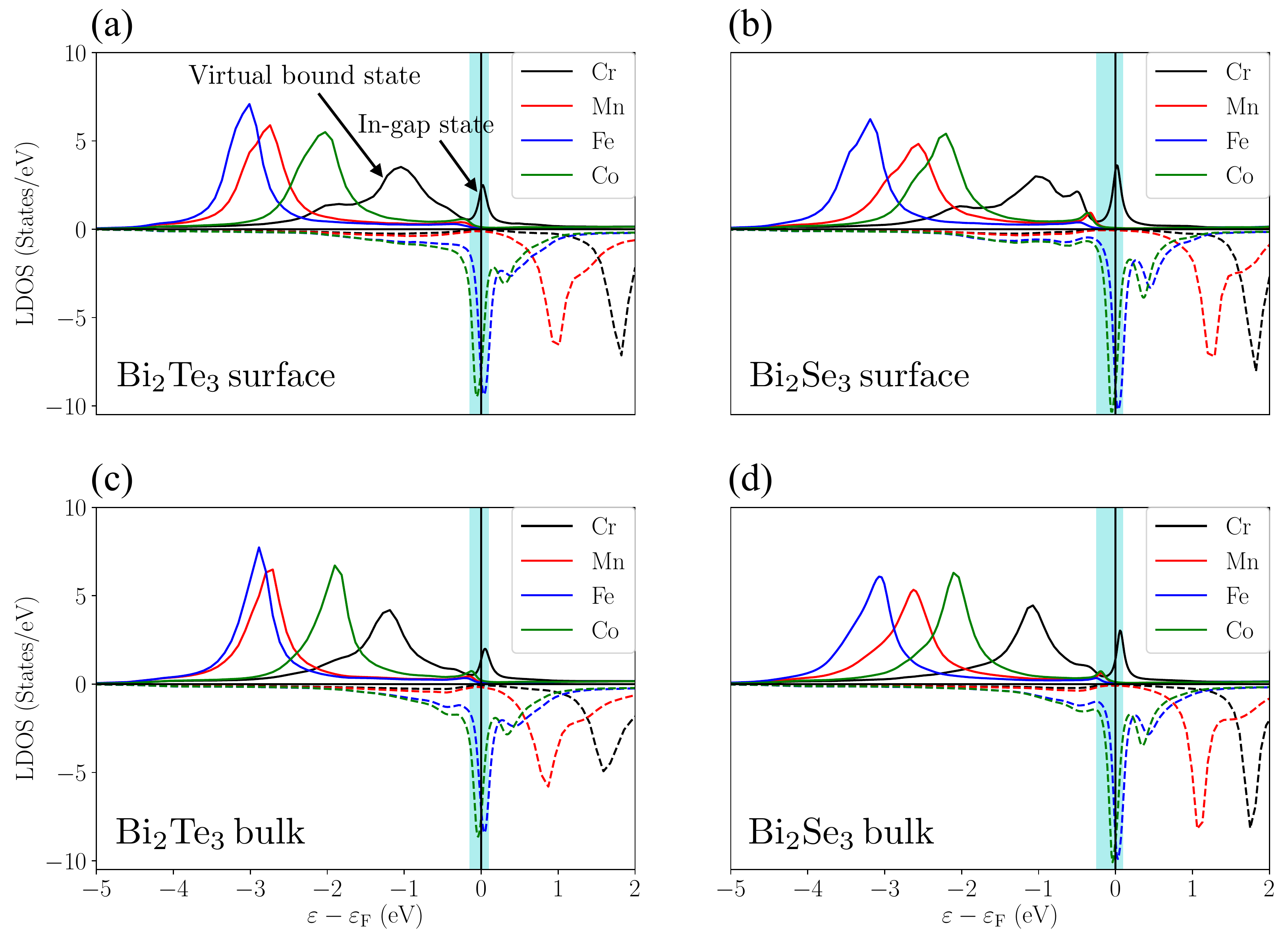}
  \caption{\label{LDOS_3d_Bi2Te3}
  First-principles simulated spin-resolved LDOS for $3d$ impurities (Cr, Mn, Fe and Co) embedded in (a) the (111) surface of Bi$_2$Te$_3$, (b) the (111) surface of Bi$_2$Se$_3$, (c) the bulk of Bi$_2$Te$_3$, and (d) in the bulk of a Bi$_2$Se$_3$ bulk.
  The majority-spin channel is represented in full lines, while the minority-spin channel is given in dashed lines.
  The energies are given with respect to $\varepsilon_\text{F}$ and the bulk band gap is highlighted in light blue.}
\end{figure*}

The LDOS of Cr, Mn, Fe and Co embedded in the Bi$_2$Te$_3$ and 
Bi$_2$Se$_3$ surfaces are plotted in Figs.~\ref{LDOS_3d_Bi2Te3}(a,b). The bulk band gap  
(light blue band in the figure) is 
$\approx 0.25$ eV for Bi$_2$Te$_3$ and $\approx 0.35$ eV for Bi$_2$Se$_3$ in agreement with 
the results of Ref.~\onlinecite{Zhang:2009}. The majority 
spin channel $(\uparrow)$ is represented in full lines, while the minority spin channel 
$(\downarrow)$ is plotted with dashed lines. On the one hand, the majority-spin channel is fully 
occupied for all considered elements with the exception of Cr. On the other hand, the minority-spin 
channel is partially occupied for Fe and Co and remains empty for Cr and Mn. 
The LDOS consists of a set of the so-called virtual bound states resulting from the 
hybridization of the atomic $d$-orbitals of the impurity with the $sp$-states of the 
Bi$_2$Te$_3$ host, resulting in a fractional valence charge of the impurity (see Table~\ref{3d_Bi2Te3_gs}). 
These resonances occur in both spin channels. Around $\varepsilon_\text{F}$, electronic states emerge 
in the bulk band gap of the substrate. These are the in-gap states central to our study, which 
were already observed for 
Cr (Mn) doped Bi$_2$Se$_3$ (Bi$_2$Te$_3$) in Ref.~\onlinecite{Chotorlishvili:2014,Antonov:2017}. 

From Fig.~\ref{LDOS_3d_Bi2Te3}, we also notice that the presence of the in-gap states 
correlates with an impurity virtual bound state being close in energy to the bulk band gap ensuring 
the presence of electrons, which can be localized at the band edges (example: majority 
spin channel of Cr, Co). However, a $d$-peak located too close to the bulk band edges 
might merge with the in-gap states (example: minority spin channel of Fe and Co in 
Bi$_2$Te$_3$). In this case, it is difficult to disentangle the in-gap state from the virtual bound state or one can even state that the latter becomes an in-gap state. For the investigated magnetic impurities, the in-gap state occurs only in one of the spin channels (either majority or minority spin), which leads to a large spin-polarization at the Fermi energy. In other words, a local half-metallic behavior emerges from the in-gap states. Our simulations also suggest that in principle Cr impurities would lead to a clear experimental observation of in-gap states because they are well separated from the virtual bound states. Of course, such an observation using scanning tunneling spectroscopy would require to consider the orbital nature of the in-gap state, which is dictated by the nature of the impurity.

The appearance of in-gap states near the bulk band edges suggests that they are connected.
In order to remove the contribution of the surface state to the in-gap states, we computed the electronic structure of the $3d$ impurities embedded in bulk Bi$_2$Te$_3$ and Bi$_2$Se$_3$.
Naturally, the impurities are substituting a Bi atom similarly to the surface case.
The change in ground state properties (charge, spin and orbital moments) of the impurities is rather small, and this can be understood from the fact that the immediate environment of the impurity (nearest neighbors) remains unchanged from surface to bulk.
The LDOS is shown in Figs.~\ref{LDOS_3d_Bi2Te3}(c,d), where we clearly observe that the in-gap state is still present for impurities in the bulk. Thus, we unambiguously prove that they originate from hybridization of the impurities $d$-states with the host bulk band edges.
In the next section, we will discuss the emergence of the in-gap states by means of a simple Anderson impurity model including contributions from a gapped bulk and a topological surface state.

\subsection{In-gap states in the Anderson model}

The results shown here are produced using the Anderson impurity model discussed in Sec.~\ref{Anderson_model_theory}.
We focus on the majority LDOS since it displays in-gap states which are not merged with the virtual bound states. Furthermore, a detailed analysis of the $lm$-resolved LDOS shows that the in-gap states are observed in the $\{d_{xy},d_{x^{2}-y^{2}},d_{xz},d_{yz}\}$ partial contributions to the LDOS, but not in the $d_{z^{2}}$ one. For clarity, we consider only the $d_{xy}$ orbital. The model parameters are obtained as follows: First, we determine the center of the band $\varepsilon^{\uparrow}_i$ of $d_{xy}$ component of the impurity LDOS.
Then, the broadening of the states is assumed to be similar for all $3d$ impurities and is used to determine the strength of the coupling to the bulk $\ev{V^\text{b}_{\vec{k}i}}$.
The occupation of the host bulk LDOS $n_\text{b}(\varepsilon)$ is also obtained from first-principles. 
The energies corresponding to the bottom of the valence and the top of the conduction are cutoffs for numerical convenience (not realistic values). A similar procedure is employed to obtain the surface state parameters.
Finally, we add an artificial broadening $\eta$ to account for the small imaginary part of the energy present in our first-principles simulations.
All the model parameters are listed in Table~\ref{Cr_Pd_Bi2Te3_model}. 

\begin{figure*}[ht]
  \centering
  \includegraphics[width=1.0\textwidth]{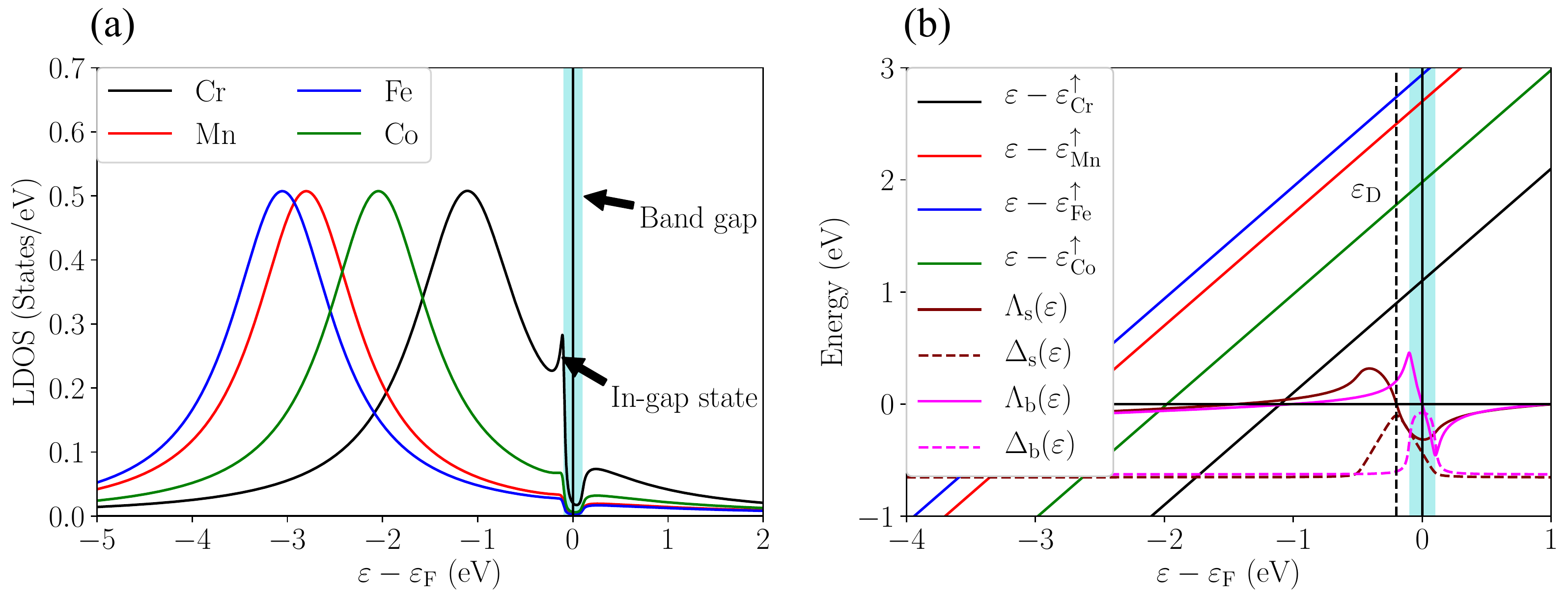}
  \caption{\label{LDOS_3d_4d_b_s_Bi2Te3_Anderson_2}
  (a) Modeled $d_{xy}$ contribution to the majority LDOS of Cr, Mn, Fe and Co embedded in Bi$_2$Te$_3$ bulk within the Anderson model.
  (b) The straight lines represent $\varepsilon - \varepsilon^{\uparrow}_{i}$, where $\varepsilon^{\uparrow}_{i}$ is the energy level of the majority $d_{xy}$-orbital.
  The real and imaginary parts of the bulk (surface) hybridization function are depicted in magenta  (brown).
  The Dirac point is located at $\varepsilon_\text{D} = -0.2$ eV.
  The used model parameters are given in Table.~\ref{Cr_Pd_Bi2Te3_model}.}
\end{figure*}

The components of the hybridization functions with the bulk and the surface state are shown in Fig~\ref{LDOS_3d_4d_b_s_Bi2Te3_Anderson_2}(b).
$\Delta_\text{b}(\varepsilon+\iu \eta)$ is a constant function in the valence and conduction bands, while it almost vanishes in the bulk band gap.
$\Lambda_\text{b}(\varepsilon+\iu \eta)$ represents the Hilbert transform of $\Delta_\text{b}(\varepsilon+\iu \eta)$ and displays sharp features at the edges of the bulk band gap.
$\Delta_\text{s}(\varepsilon+\iu \eta)$ has a linear behavior for 
$\varepsilon \in [-0.5,0.1]\,$eV and is a constant otherwise.
$\Lambda_\text{s}(\varepsilon+\iu \eta)$ does not have any sharp feature.
In Fig~\ref{LDOS_3d_4d_b_s_Bi2Te3_Anderson_2}(a), we show the model majority LDOS for our $3d$ impurities embedded in Bi$_2$Te$_3$ bulk.
The in-gap states observed at the bulk band edges emerge due to high values in $\Lambda_\text{b}(\varepsilon+\iu \eta)$ combined with  small values for $\Delta_\text{b}(\varepsilon+\iu \eta)$ (gap region).
The model reproduces qualitatively the position and shape of the in-gap states for Mn, Fe and Co.
Furthermore, we obtain a clear feature in the LDOS of Cr. However, it is located at the lower bulk 
band edge in contrast to what is observed from first-principles as shown in Fig.~\ref{LDOS_3d_Bi2Te3}(c).

\begin{table}
\begin{center}
\begin{tabular}{|l|l|l|l|l|l|l|l|}
\hline
Bi$_2$Te$_3$ (bulk) & $\varepsilon_\text{bv}$  &  $\varepsilon_\text{tv}$   &  $\varepsilon_\text{bc}$   &   $\varepsilon_\text{tc}$  &  $n_\text{b}$   &  $\ev{V^\text{b}_{\vec{k}i}}$  \\
\hline
- - -   &  -10.0  &  -0.1   &   0.1  & 10.0  &  0.15   &  1.15    \\
\hline
Bi$_2$Te$_3$ (surface) & $\varepsilon_\text{bv}$  &   $\varepsilon_\text{tc}$  &  $\varepsilon_{0}$   & $\varepsilon_\text{D}$ &  $\ev{V^\text{s}_{\vec{k}i}}$ & - - -  \\
\hline
- - - &  -10.0  & 10.0  &  0.3  & -0.2  & 0.25 & - - - \\
\hline
$3d$ elements & $\varepsilon^{\uparrow}_\text{Cr}$ & $\varepsilon^{\uparrow}_\text{Mn}$ &  $\varepsilon^{\uparrow}_\text{Fe}$ & $\varepsilon^{\uparrow}_\text{Co}$ & $\eta$ & - - -  \\
\hline
- - -   & -1.10 & -2.70 & -2.94 & -1.98 & 0.02 & - - - \\
\hline
\end{tabular}
\caption{Anderson model parameters used to compute the majority LDOS for the considered $3d$ impurities, the bulk hybridization function and surface hybridization function. $\eta$ is an artificial broadening added to mimic the small imaginary part of the energy included in our first-principles simulations. All the parameters are given in eV except $n_\text{b}$ which is given in states/eV.}
\label{Cr_Pd_Bi2Te3_model}
\end{center}
\end{table}

Considering only the topological surface state in Eq.~\eqref{ldos_model_anderson}, the condition to observe an in-gap state is: $\varepsilon - \varepsilon^{\uparrow}_{i} \simeq \Lambda_\text{s}(\varepsilon)$ and $\Delta_\text{s}(\varepsilon)$ must be small in order to increase the spectral weight. In other words,  $\Lambda_\text{s}(\varepsilon)$ and $\varepsilon - \varepsilon^{\uparrow}_{i}$ must ideally cross each other in a region where the substrate LDOS is low, \textit{i.e}. close to the Dirac point (near $\varepsilon_\text{D}$).
As an example, we take $\ev{V^\text{s}_{\vec{k}i}} = 0.25$ eV 
(smaller compared to the bulk one, see Table \ref{Cr_Pd_Bi2Te3_model}) and plot in Fig.~\ref{LDOS_3d_4d_b_s_Bi2Te3_Anderson_2} the real and imaginary parts of ${\Sigma}^\text{s}_{i}(\varepsilon)$ and $\varepsilon - \varepsilon^{\uparrow}_{i}$ for the considered $3d$ impurities.
The crossing near $\varepsilon_\text{D}$ leading to an in-gap state 
is not observed for this particular case.
Although it may occur for stronger couplings to the surface, this would be 
in contradiction with our first-principles calculations 
predicting a weak coupling to the surface state.

\section{Conclusions}
\label{conclusions_in_gap_states}
In this paper, we combined a first-principles and model approach to understand the 
emergence and origins of in-gap states in the LDOS of $3d$ transition metal impurities 
embedded in topologically insulating hosts, which consist of Bi$_2$Te$_3$ and 
Bi$_2$Se$_3$. We considered bulk systems and thin films. We found that ground 
state properties such as the valence charge, spin and orbital moment on the impurity 
can be affected when trading the Bi$_2$Te$_3$ host for the Bi$_2$Se$_3$ 
one. The largest changes were noticed for the orbital moments. 
Our first-principles simulations showed the emergence of in-gap states 
when the impurities are embedded in the bulk, ruling out the necessity of the 
topological surface state for their creation. Furthermore, we built an Anderson 
model where the impurity contains a single $d$-orbital which hybridizes with 
bulk and surface states. Within this model, we showed that the in-gap states 
arise at the bulk band edges from the real part of the bulk hybridization function.
We also considered the possibility of creating in-gap states when considering 
solely the topological surface state. However, this requires large and nonphysical 
coupling constants between the $3d$ impurities and the topological surface states. 

For the investigated systems, the in-gap states 
are found in one single spin-channel, which generates a half-metallic behavior at the impurity site 
and its immediate surrounding. Also the orbital nature of these localized states depend on 
the electronic filling of the impurity.

Our results are in good agreement with the STS measurements 
obtained in Refs.~\onlinecite{Honolka:2012,Sessi:2016}, which display a finite local 
density of states within the gap region when the tip is located above the defects 
($3d$ impurities). This finite density of states is attributed to the presence of 
in-gap states leading to a filling of the band gap locally. Furthermore, the in-gap 
states reported in this work were also observed previously from first-principles 
calculations for Cr (Mn) in Bi$_2$Se(Te)$_3$ in Refs.~\onlinecite{Chotorlishvili:2014,Antonov:2017} but without relating their existence to the hybridization with the bulk bands. 
We point out that although the investigated single magnetic impurities do not open 
a band gap locally, this does not exclude the existence of a quantum 
anomalous Hall state~\cite{Sessi:2016,Peixoto:2016}.

Finally, the in-gap states provide a relatively high density of states 
at the Fermi energy which may profoundly alter the magnetic properties 
of the system, for instance: the magnetic anisotropy energy~\cite{Pick:2003,Wang:1993}, 
the response of the impurities to external time-dependent perturbations~\cite{Samir:2010,mdsd:2015,Samir:2015}, 
their magnetic stability against spin fluctuations~\cite{Julen:2016} and 
many other phenomena. These properties are currently under investigation.  

\textbf{Acknowledgements} We thank P. R\"ussmann for fruitful discussions and for providing the potentials 
and initial setups for the topological insulating host (Bi$_2$Te$_3$). This work 
was supported by the European Research Council (ERC) under the European Union's Horizon 
2020 research and innovation programme (ERC-consolidator grant 681405 DYNASORE).
We gratefully acknowledge the computing time granted by the JARA-HPC Vergabegremium 
and VSR commission on the supercomputer JURECA at Forschungszentrum J\"ulich.

\appendix
\section{Bulk and surface hybridization function for the Anderson model}
\label{Append_A}
Here, we analytically derive the real and imaginary parts of $\Sigma^\text{b}_{i}(\varepsilon)$ and $\Sigma^\text{b}_{i}(\varepsilon)$. We assume that $V^\text{b}_{\vec{k}i}$ is weakly depending on $\vec{k}$:
\begin{equation}
\label{hybridization_funct_bulk_2}
\begin{split}
\Sigma^\text{b}_{i}(\varepsilon+\iu \eta) & = \sum_{\vec{k}} \frac{|V^\text{b}_{\vec{k}i}|^{2}}{\varepsilon - 
	\varepsilon_{\vec{k}}+\iu \eta}\quad,\\
& = |\ev{V^\text{b}_{\vec{k}i}}|^{2} \sum_{\vec{k}} \frac{1}{\varepsilon - 
	\varepsilon_{\vec{k}}+\iu \eta}\quad,\\
& = |\ev{V^\text{b}_{\vec{k}i}}|^{2} \int_{}^{} d\varepsilon^{\prime} \frac{n_\text{b}(\varepsilon^\prime)}{\varepsilon-\varepsilon^{\prime}+\iu \eta} \quad.
\end{split}
\end{equation} 

Using the definition of $n_\text{b}(\varepsilon)$ given in Eq.~\eqref{bulk_ldos_and}, $\Lambda_\text{b}(\varepsilon+\iu \eta)$ reads: 
\begin{equation}
\begin{split}
\Lambda_\text{b}(\varepsilon+\iu \eta)  = &-\frac{n_\text{b}|\ev{V^\text{b}_{\vec{k}i}}|^{2}}{2}\,\left[\ln\left(\frac{(\varepsilon-\varepsilon_\text{tv})^{2}
	+\eta^2}{(\varepsilon-\varepsilon_\text{bv})^{2}+\eta^2}\right)\right.\\
&+\left.\ln\left(\frac{(\varepsilon-\varepsilon_\text{tc})^{2}
	+\eta^2}{(\varepsilon-\varepsilon_\text{bc})^{2}+\eta^2}\right) 
\right]\quad,
\end{split}
\end{equation}
and the imaginary part is:
\begin{equation}
\begin{split}
\Delta_\text{b}(\varepsilon+\iu \eta) &  = n_\text{b}|\ev{V^\text{b}_{\vec{k}i}}|^{2} \,\left[\arctan\left(\frac{\varepsilon-\varepsilon_\text{tv}}{\eta}\right)-\arctan\left(\frac{\varepsilon-\varepsilon_\text{bv}}{\eta}\right)\right] \\
& + n_\text{b}|\ev{V^\text{b}_{\vec{k}i}}|^{2}\,\left[\arctan\left(\frac{\varepsilon-\varepsilon_\text{tc}}{\eta}\right)
-\arctan\left(\frac{\varepsilon-\varepsilon_\text{bc}}{\eta}\right)\right]\quad.
\end{split}
\end{equation}
Once more, assuming that $V^\text{s}_{\vec{k}i}$ depends weakly on $\vec{k}$, the real part of ${\Sigma}^\text{s}_{i}(\varepsilon)$ reads:
\begin{widetext}
	\begin{equation}
	\begin{split}
	& \Lambda_\text{s}(\varepsilon+\iu \eta) =  \frac{|\ev{V^\text{s}_{\vec{k}i}}|^{2}\varepsilon}{2\varepsilon^{2}_{0}}\left[\ln\left(\frac{\varepsilon_\text{r}^{2}+\eta^2}{(\varepsilon_\text{r}+\varepsilon_{0})^{2}+\eta^2}\right)-\ln\left(\frac{(\varepsilon_\text{r}-\varepsilon_{0})^{2}+\eta^2}{\varepsilon_\text{r}^{2}+\eta^2}\right)\right]  \\&
	+ \frac{|\ev{V^\text{s}_{\vec{k}i}}|^{2}\eta}{\varepsilon^{2}_{0}}\, \left[2\arctan\left(\frac{\varepsilon_\text{r}}{\eta}\right)-\arctan\left(\frac{\varepsilon_\text{r}+\varepsilon_{0}}{\eta}\right)-\arctan\left(\frac{\varepsilon_\text{r}-\varepsilon_{0}}{\eta}\right)\right] \\&
	-\frac{|\ev{V^\text{s}_{\vec{k}i}}|^{2}}{2\varepsilon_{0}}\,\left[\ln\left(\frac{(\varepsilon_\text{r}+\varepsilon_0)^{2}+\eta^2}{(\varepsilon_\text{r}-\varepsilon_\text{bv})^{2}+\eta^2}\right)+\ln\left(\frac{(\varepsilon_\text{r}-\varepsilon_\text{tc})^{2}+\eta^2}{(\varepsilon_\text{r}-\varepsilon_0)^{2}+\eta^2}\right) \right]\quad
	\end{split}
	\label{Ti_Green_func_surface_hydirdization function}
	\end{equation}
\end{widetext}
while the imaginary part reads: 
\begin{widetext}
	\begin{equation}
	\begin{split}
	& \Delta_\text{s}(\varepsilon+\iu \eta) =  \frac{|\ev{V^\text{s}_{\vec{k}i}}|^{2}\eta}{2\varepsilon^{2}_{0}}\left[\ln\left(\frac{\varepsilon_\text{r}^{2}+\eta^2}{(\varepsilon_\text{r}+\varepsilon_{0})^{2}+\eta^2}\right)-\ln\left(\frac{(\varepsilon_\text{r}-\varepsilon_{0})^{2}+\eta^2}{\varepsilon_\text{r}^{2}+\eta^2}\right)\right]  \\&
	+ \frac{|\ev{V^\text{s}_{\vec{k}i}}|^{2}\varepsilon_\text{r}}{\varepsilon^{2}_{0}}\, \left[2\arctan\left(\frac{\varepsilon_\text{r}}{\eta}\right)-\arctan\left(\frac{\varepsilon_\text{r}+\varepsilon_{0}}{\eta}\right)-\arctan\left(\frac{\varepsilon_\text{r}-\varepsilon_{0}}{\eta}\right)\right] \\&
	+ \frac{|\ev{V^\text{s}_{\vec{k}i}}|^{2}}{\varepsilon_{0}} \,\left[\arctan\left(\frac{\varepsilon_\text{r}+\varepsilon_0}{\eta}\right)-\arctan\left(\frac{\varepsilon_\text{r}-\varepsilon_\text{bv}}{\eta}\right)\right] \\
	& + \frac{|\ev{V^\text{s}_{\vec{k}i}}|^{2}}{\varepsilon_{0}}\,\left[\arctan\left(\frac{\varepsilon_\text{r}-\varepsilon_\text{tc}}{\eta}\right)-\arctan\left(\frac{\varepsilon_\text{r}-\varepsilon_0}{\eta}\right)\right]\quad,
	\end{split}
	\label{Ti_Green_func_surface_hydirdization function_2}
	\end{equation} 
\end{widetext}
with $\varepsilon_\text{r} = \varepsilon - \varepsilon_\text{D}$. 

\bibliography{mylib_paper.bib}{}

\end{document}